\documentclass[a4paper,11pt]{article}
\usepackage{graphicx,amsmath,amssymb,wrapfig}
\usepackage{pos}
\usepackage{slashed}
\usepackage{color}

\definecolor{mycolor}{rgb}{0.6,0.0,0.4}

\definecolor{mycolorg}{rgb}{0.0,0.6,0.2}

\title{Possible studies on generalized parton distributions\\
       and gravitational form factors in neutrino reactions}
\ShortTitle{Possible studies on GPDs and gravitational form factors in neutrino reactions}

\author*[a,b]{S. Kumano}
\author[c]{R. Petti}

\affiliation[a]
  {KEK Theory Center, Institute of Particle and Nuclear Studies, KEK,\\
   Oho 1-1, Tsukuba, Ibaraki, 305-0801, Japan}
\affiliation[b]
  {J-PARC Branch, KEK Theory Center,
   Institute of Particle and Nuclear Studies, KEK,\\
   and Theory Group, Particle and Nuclear Physics Division, J-PARC Center,\\
   203-1, Shirakata, Tokai, Ibaraki, 319-1106, Japan}
\affiliation[c]
  {Department of Physics and Astronomy, University of South Carolina,\\
   Columbia, South Carolina, 29208, USA}
\emailAdd{shunzo.kumano@kek.jp}
\emailAdd{Roberto.Petti@cern.ch}

\abstract{
Spacelike and timelike generalized parton distributions (GPDs)
have been investigated in charged-lepton scattering and electron-positron collisions
via deeply virtual Compton scattering and two-photon processes, respectively.
Furthermore, we expect that hadron-accelerator-facility measurements will 
be performed in future. The GPDs will play a crucial role in clarifying 
the origins of hadron spins and masses in terms of quarks and gluons. 
It is also possible to probe internal pressure within hadrons 
for understanding their stability.
Gravitational form factors of hadrons used to be considered as
a purely academic subject because gravitational interactions
are too weak to be measured in microscopic systems.
However, due to the development of hadron-tomography field,
it became possible to extract the gravitational form factors 
from the actual GPD measurements without relying on direct
gravitational interactions.
Neutrino reactions can also be used for GPD studies in future, for example, 
by using the Long-Baseline Neutrino Facility at Fermilab. 
The neutrino GPD measurements are valuable especially for finding 
the flavor dependence of the GPDs in a complementary way to
the charged-lepton experiments. We give an overview of the GPDs
and discuss possible neutrino GPD measurements using
the single-pion production processes
$\nu + N \to \ell^- + N' + \pi$ 
and $\bar\nu + N \to \ell^+ + N' + \pi$.
}

\begin{document}
\maketitle

\section{Introduction}
\label{introduction}

Unpolarized structure functions $F_2$ and $F_3$ of the nucleon
were measured by neutrino deep inelastic inelastic scattering (DIS)
from heavy nuclei with appropriate nuclear corrections \cite{nutev}.
These structure functions are expressed by 
collinear parton distribution functions (PDFs), which indicate
longitudinal momentum distributions of partons.
In recent years, three-dimensional structure functions have been
investigated extensively for clarifying the transverse structure
of the nucleon in addition to the longitudinal distributions
and for understanding the origin of the nucleon spin 
including the partonic orbital-angular-momentum (OAM) contribution.
The OAM contribution should be determined by generalized
parton distributions (GPDs) \cite{Diehl:2003ny},
which have been measured by deeply virtual Compton scattering (DVCS)
and meson productions at charged-lepton accelerator facilities
in the spacelike region.
The DVCS has been investigated 
by the HERMES and COMPASS collaborations and also 
at the Thomas Jefferson National Accelerator Facility (JLab).
In 2030's, it will be investigated at electron-ion colliders
in US and China (EIC, EicC) \cite{eic-nu}.
There are also possibilities of measuring the GPDs at hadron
facilities by using high-energy exclusive reactions
such as at the Japan Proton Accelerator Research Complex (J-PARC)
\cite{hadron-gpds}. 
All of these are spacelike GPD studies, whereas it is possible
to investigate the timelike GPDs \cite{Kumano:2017lhr}, 
which are also called generalized distribution amplitudes (GDAs), 
by two-photon processes in $e^+ e^-$ annihilation, 
for example, at the KEK-B factory.

There is another important purpose to investigate the GPDs
for understanding hadron masses and their internal pressures 
in terms of quark and gluon degrees of freedom \cite{Kumano:2017lhr}.
The studies on the origin of hadron masses by hadron-mass decomposition 
are now becoming one of major purposes for building the future EICs. 
The GPD measurements have been done mainly at charged-lepton accelerator
facilities, 
and there is no GPD measurement in neutrino reactions at this stage.
However, we may recollect that the neutrino DIS experiments have been important 
in determining the unpolarized PDFs, especially on 
the strange-quark distribution via the opposite-sign dimuon events
and valence-quark distributions via the structure function $F_3$.
Considering these past experiences, we expect that 
future neutrino experiments could provide valuable information
on the GPDs in a complementary way 
to the charged-lepton and hadron-facility measurements \cite{eic-nu}.
The Long-Baseline Neutrino Facility (LBNF) at Fermilab
can supply (anti)neutrino beams in the energy region of 2-15 GeV
\cite{lbnf-beam} allowing to measure the GPDs, for example, 
by the pion-production reaction 
$\nu_\mu + N \to \mu +\pi+N'$
\cite{neutrino-GPDs,psw-2017,Kumano:2018bwh}.
In general, (anti)neutrino Charge Current (CC) interactions 
are sensitive to the quark flavor, offering a valuable tool 
to study the flavor dependence of the GPDs together with charged-lepton data, 
and to investigate the origin of hadron spins and masses.
In this article, we discuss such a possibility.

\section{Generalized parton distribution functions}
\label{GPDs}

\begin{wrapfigure}[17]{r}{0.34\textwidth}
\vspace{-0.70cm}
\begin{center}
    \includegraphics[width=5.0cm]{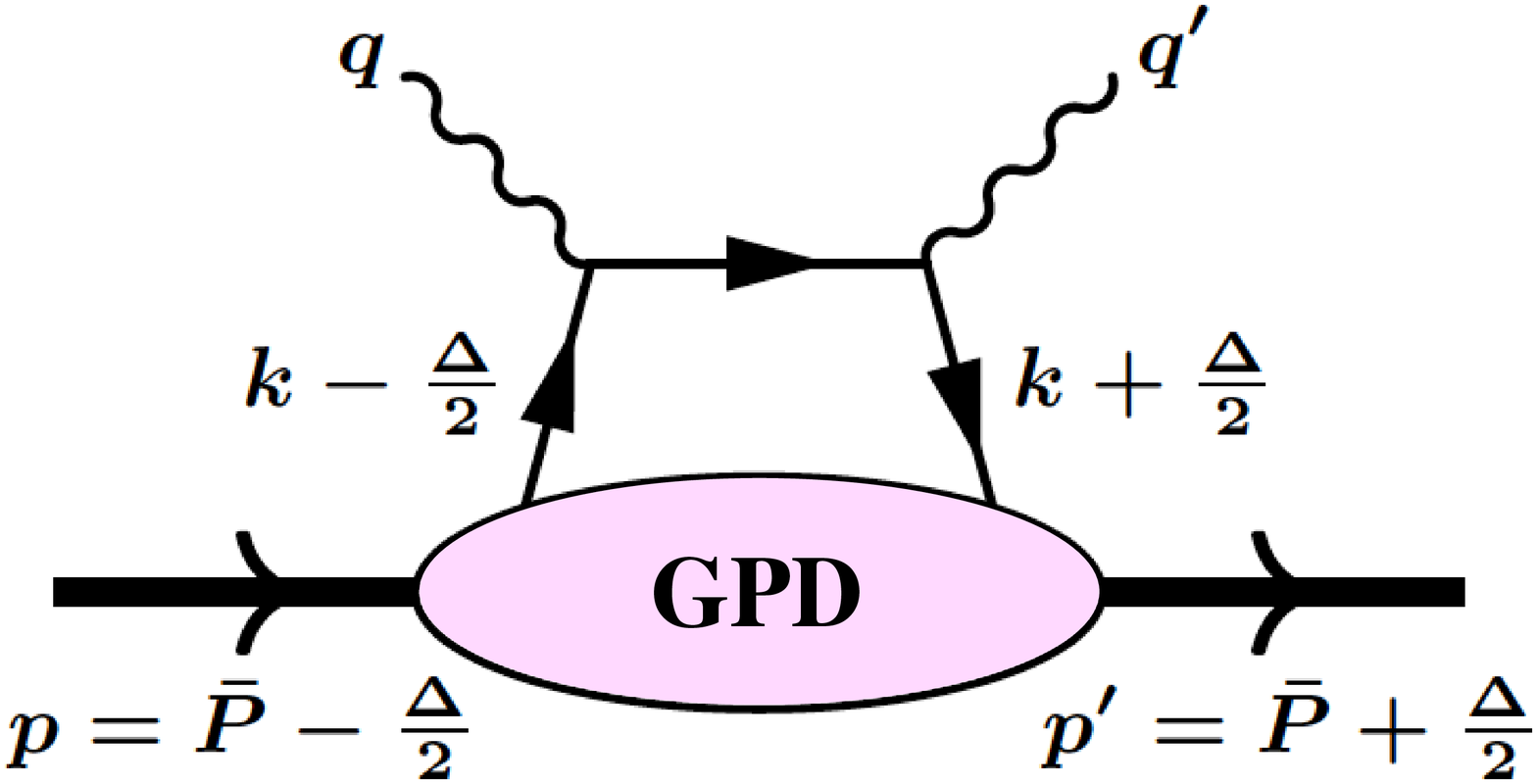}
\vspace{-0.20cm}
\caption{Spacelike GPDs in virtual Compton scattering.}
\label{fig:DVCS-GPD}
\end{center}
\vspace{-0.80cm}
\begin{center}
    \includegraphics[width=4.5cm]{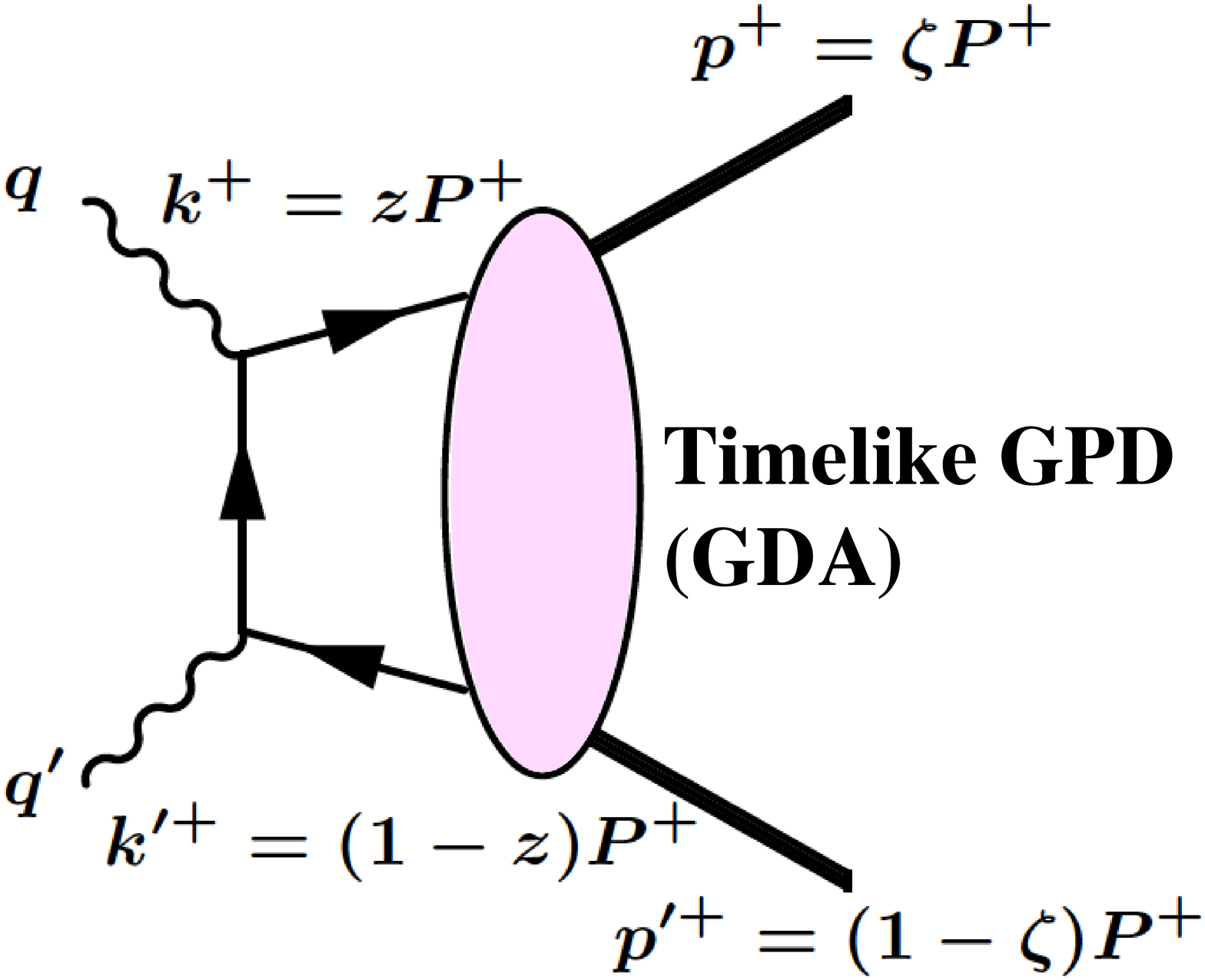}
\vspace{-0.20cm}
\caption{Timelike GPDs (GDAs) in two-photon process.}
\label{fig:2gamma-GPD}
\vspace{-0.5cm}
\end{center}
\end{wrapfigure}

The spacelike GPDs of the nucleon are measured, for example, 
by the deeply virtual Compton scattering (DVCS) 
at charged-lepton accelerator facilities as shown 
in Fig.\,\ref{fig:DVCS-GPD}. The photon momenta 
are $q$ and $q'$, and the nucleon momenta are $p$ and $p'$.
We define average momenta ($\bar P$, $\bar q$)
and momentum transfer $\Delta$ as 
$\bar P = (p+p')/2$,
$\bar q = (q+q')/2$, and
$\Delta = p'-p = q-q'$.
Three variables for expressing the GPDs are 
the Bjorken variable $x$, the skewness parameter $\xi$,
and the momentum-transfer squared $t$ are defined by
$x = Q^2/(2p \cdot q)$, 
$\xi = \bar Q^2/(2 \bar P \cdot \bar q)$, and
$t = \Delta^2$.
By the lightcone momentum notations, $x$ and $\xi$ are expressed as
$x=k^+ / P^+$ and $\xi = -\Delta^+ / (2P^+)$ with $P=p+p'$.
If the kinematical condition $Q^2 \gg |t|,\ \Lambda_{\text{QCD}}^2$,
where $\Lambda_{\text{QCD}}$is the QCD scale parameter, 
is satisfied, the DVCS process is factorized 
into the hard part and the soft one by the GPDs 
$H^q$ and $E^q$ defined in the matrix element
\begin{align}
\!
 \int & \frac{d y^-}{4\pi}e^{i x P^+ y^-}
 \! \!
 \left< p' \left| 
 \bar{q}(-y/2) \gamma^+ q(y/2) 
 \right| p \right> 
  _{y^+ = \vec y_\perp =0}
\! =  \!
 \frac{1}{2  P^+} \bar{u} (p') 
 \! \left [ H^q (x,\xi,t) \gamma^+
   \! \!  +  \!
     E^q (x,\xi,t)  \frac{i \sigma^{+ \alpha} \Delta_\alpha}{2 m_N}
 \right ] \! u (p) .
\label{eqn:gpd-vector}
\nonumber\\[-0.35cm]
\end{align}
\ \vspace{-0.85cm}

\noindent
In pion-production and neutrino cross sections,
there are other GPDs $\tilde{H}^q $ and $\tilde{E}^q$ 
associated with 
the matrix element of 
the axial-vector current as
\begin{align}
\! \! \!
 \int & \frac{d y^-}{4\pi}e^{i x P^+ y^-}
 \! \!
 \left< p' \left| 
 \bar{q}(-y/2) \gamma^+ \gamma_5 q(y/2) 
 \right| p \right> 
 _{y^+ = \vec y_\perp =0}
\! =  \! \frac{1}{2  P^+} \bar{u} (p') 
\! \left [ \tilde{H}^q (x,\xi,t) \gamma^+ \gamma_5
    \! + \! \tilde{E}^q (x,\xi,t)  \frac{\gamma_5 \Delta^+}{2 m_N}
 \right ] \! u (p) .
\label{eqn:gpd-axial-v}
\nonumber\\[-0.20cm]
\end{align}
\ \vspace{-1.10cm}

The unique features of the GPDs are that
they become the unpolarized and longitudinally-polarized PDFs
in the forward limit:
$H^q (x, 0, 0) = q(x)$, $\tilde{H}^q (x, 0, 0) = \Delta q(x)$,
that their first moments are the corresponding form factors:
$\int_{-1}^{1} dx H^q(x,\xi,t) = F_1^q (t)$,
$\int_{-1}^{1} dx E^q(x,\xi,t) = F_2^q (t)$, 
$\int_{-1}^{1} dx \tilde{H}^q(x,\xi,t) = g_A^q (t)$, 
$\int_{-1}^{1} dx \tilde{E}^q(x,\xi,t) = g_P^q (t)$,
and that the second moment is the quark contribution to the nucleon spin:
$J_q  =  \int dx \, x \, [ H^q (x,\xi,t=0) +E^q (x,\xi,t=0) ] /2
      = \Delta q^+ /2 + L_q$.
Here, $L_q$ is a quark orbital-angular-momentum contribution ($L_q$) 
to the nucleon spin. Since we know the quark-spin contribution $\Delta q^+$
from experimental measurements, it is possible to determine $L_q$
from the GPD measurements. 

The timelike GPDs are often called the GDAs,
and they are measured by the $s$-$t$ crossed process of the DVCS, 
so called the two-photon process, as shown in Fig.\,\ref{fig:2gamma-GPD}. 
The GDAs  or timelike GPDs are defined by the matrix element similar
to Eqs.\,(\ref{eqn:gpd-vector}) and (\ref{eqn:gpd-axial-v})
between the vacuum and the final hadron pair $h \bar h$
\cite{Kumano:2017lhr}.
For example, they are defined for the $\pi^0$ pair as
\begin{align}
& \Phi_q^{\, \pi^0 \pi^0} (z,\zeta,W^2) 
  = \int \frac{d y^-}{2\pi}\, e^{i (2z-1)\, P^+ y^- /2}
  \langle \, \pi^0 (p) \, \pi^0 (p') \, | \, 
 \bar{q}(-y/2) \gamma^+ q(y/2) 
  \, | \, 0 \, \rangle \Big |_{y^+=\vec y_\perp =0} \, .
\label{eqn:gda-pi}
\end{align}
The GDAs are expressed by three variables,
the momentum fractions $z$ and $\zeta$ in Fig.\,\ref{fig:2gamma-GPD}
and the invariant-mass squared $W^2$ as
$z = k^+ / P^+$,
$\zeta = p^+ / P^+ = (1+\beta \cos\theta)/2$,
and $W^2 =s$, where $\beta$ is defined by
$\beta =|\vec p \,|/p^0 = \sqrt{1-4m_\pi^2/W^2}$,
and $\theta$ is the scattering angle in the center-of-mass frame
of the final pions.
The two-photon process is factorized if the condition 
$Q^2 \gg W^2,\ \Lambda_{\text{QCD}}^2$ is satisfied
to express it in terms of the GDAs.
The corresponding spacelike GPDs for the pion are given as
\begin{align}
H_q^{\, \pi^0} (x,\xi,t)
=  \int\frac{d y^-}{4\pi} \, e^{i x \bar P^+ y^-}
\! \left< \pi^0 (p') \left| 
\bar{q}(-y/2) \gamma^+ q(y/2) 
 \right| \pi^0 (p) \right> \Big |_{y^+ = \vec y_\perp =0} .
\label{eqn:gpd-pi}
\end{align}
The spacelike and timelike GPDs are related with each other 
by the $s$-$t$ crossing as 
$\Phi_q^{\,\pi^0 \pi^0} (z',\zeta,W^2)$
$ \leftrightarrow
   H_q^{\,\pi^0} \left ( x=(1-2z')/(1-2\zeta),
            \, \xi=1/(1-2\zeta), \, t=W^2 \right )$.

\section{Gravitational form factors of hadrons from spacelike and timelike GPDs}
\label{GFFs}

\begin{wrapfigure}[8]{r}{0.50\textwidth}
\vspace{-0.80cm}
\begin{center}
    \includegraphics[width=7.6cm]{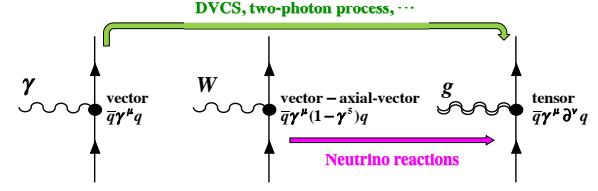}
\vspace{-0.60cm}
\caption{Gravitational form factors from electromagnetic
and weak interactions.}
\label{fig:grav-em-weak}
\end{center}
\end{wrapfigure}

Electromagnetic and weak form factors of hadrons and nuclei
have been measured in lepton scatterings, whereas gravitational
form factors used to be considered purely theoretical quantities until
recently, because gravitational interactions are too weak
to be used for measuring interactions with microscopic particles.
However, it became possible to measure them without 
direct gravitational interactions by hadron tomography
techniques as illustrated in Fig.\,\ref{fig:grav-em-weak}.
In order to understand why the gravitational form factors
can be obtained from electromagnetic and weak interactions,
we consider moments of the nonlocal operators 
in Eqs.\,(\ref{eqn:gpd-vector}), (\ref{eqn:gpd-axial-v}),
(\ref{eqn:gda-pi}), and (\ref{eqn:gpd-pi}) as
\begin{align}
\left ( \frac{P^+}{2}  \right ) ^n
\int dx \, x^{\, n-1} 
\! \! \int\frac{d y^-}{2\pi}e^{i x P^+ y^- /2} \,
  \bar{q}(-y/2) \gamma^+ q(y/2) \Big |_{y^+ = \vec y_\perp =0}
 = \bar q (0) \gamma^+ 
  \left ( i \overleftrightarrow \partial^+  \right )^{n-1} \! q(0) ,
\label{eqn:tensor-int-gpd}
\end{align}
where $\overleftrightarrow \partial$ is defined by
$f_1 \overleftrightarrow \partial f_2 
 = [ f_1 (\partial f_2)  - (\partial f_1) f_2 ]/2$.
For $n=1$, it is the ordinary vector-type electromagnetic current;
however, we notice that the operator is the energy-momentum tensor 
of a quark for $n=2$ \cite{Kumano:2017lhr}.
It indicates that the GPDs contain the information on 
the gravitational form factors.
Therefore, the second moments of the spacelike and timelike GPDs 
are given by the matrix elements of the energy-momentum tensor 
$T_q^{\,\mu\nu}$, and they are expressed by the spacelike 
and timelike gravitational form factors $\Theta_1$ 
and $\Theta_2$ as
\begin{align}
\int_{-1}^1  dx \, x \, 
H_q^{\, \pi^0} (x,\xi,t)
& 
= \frac{1}{(P^+)^2} \langle \, \pi^0 (p') \, | \, T_q^{++} (0) \,
       | \,  \pi^0 (p) \, \rangle   
\nonumber \\[-0.10cm]
&
= \frac{1}{2 \, {(P^+)^2}} 
  \left [ \, \left ( t \, g^{++} -q^{+} q^+ \right ) \, \Theta_{1, q} (t)
                + P^{+} P^+ \,  \Theta_{2, q} (t) \,
  \right ] ,
\\
\int_0^1  dz \, (2z -1) \, 
\Phi_q^{\pi^0 \pi^0} (z,\,\zeta,\,W^2) 
& = \frac{2}{(P^+)^2} \langle \, \pi^0 (p) \, \pi^0 (p') \, | \, T_q^{++} (0) \,
       | \, 0 \, \rangle 
\nonumber \\[-0.10cm]
&
= \frac{1}{(P^+)^2} 
  \left [ \, \left ( t \, g^{++} -q^{+} q^+ \right ) \, \Theta_{1, q} (t)
                + P^{+} P^+ \,  \Theta_{2, q} (t) \,
  \right ] .
\label{eqn:integral-over-xz}
\end{align}
Here, the energy-momentum tensor is given by
$ T_q^{\,\mu\nu} (x) = \overline q (x) \, \gamma^{\,(\,\mu} 
   i \overleftrightarrow D^{\nu)} \, q (x)$,
with the convention 
$A^{(\mu} B^{\nu)} = ( A^{\mu} B^{\nu} + A^{\nu} B^{\mu} )/2$ and
the covariant derivative 
$D^\mu = \partial^{\,\mu} -ig \lambda^a A^{a,\mu}/2$
with the QCD coupling constant $g$ 
and the SU(3) Gell-Mann matrix $\lambda^a$.
For the spin-1/2 nucleons, the spacelike GPDs are related to the gravitational
form factors $A_q$, $B_q$, $C_q$, and $\bar C_q$ in the same way as
\begin{align}
& 
\bar u (p') 
\left [ \, 
   \int_{-1}^1  dx \, x \, H_q (x,\xi,t) \, \gamma^+ 
+  \int_{-1}^1  dx \, x \, E_q (x,\xi,t) \,
       \frac{i \sigma^{+\sigma} \Delta_\sigma}{2 M_N}
\, \right ] u(p) 
= \frac{1}{\bar P^+} \langle \, N (p') \, | \, T_q^{++} (0) \, | \, N (p) \, \rangle   
\nonumber \\
& \ \ 
= \frac{1}{\bar P^+} \bar u (p')
  \left [ \, A_q (t) \gamma^{+} \bar P^{+}  
  + B_q (t) \frac{\bar P^{+} i \sigma^{+\sigma}\Delta_\sigma}{2 M_N}
  + D_q (t) \frac{\Delta^+ \Delta^+ -g^{++} \Delta^2}{M_N}
  + \bar C_q (t) M_N g^{++}
  \, \right ] u(p) .
\label{eqn:nucleon-gpd}
\end{align}
Therefore, these various GPD measurements enable extraction
of the gravitational form factors of hadrons \cite{Kumano:2017lhr}
and also clarifications of the origins of hadron spins and masses.

\vfill\eject

\section{GPDs in neutrino reactions}
\label{GPDs-in-nu}

\subsection{Pion-production cross section and GPDs in neutrino reactions} 
\label{neutrino-cross} 

\begin{wrapfigure}[8]{r}{0.36\textwidth}
\vspace{-0.10cm}
\begin{center}
\begin{minipage}[c]{0.36\textwidth}
    \vspace{-0.70cm}\hspace{+0.37cm}
    \includegraphics[width=4.0cm]{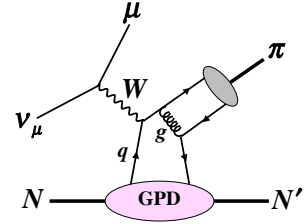}
\vspace{-0.20cm}
\caption{GPDs in neutrino scattering.}
\label{fig:nu-GPD}
\vspace{-0.5cm}
\end{minipage}
\end{center}
\end{wrapfigure}

There have been a number of works on the GPDs in neutrino reactions
\cite{neutrino-GPDs,psw-2017}. 
Instead of introducing all of these results,
we explain the formalism and numerical results
by Pire, Szymanowski, and Wagner
\cite{psw-2017} as a recent work in the following.
Instead of the virtual-Compton-like process, it is appropriate
to rely on larger meson-production cross sections in neutrino reactions.
For example, the pion production process $\nu N \to \ell^- N'\pi$
is expressed by a typical subprocess with the GPDs 
in Fig.\,\ref{fig:nu-GPD}. In the pion production,
quark and gluon GPD contributions to the amplitude 
are generally given as \cite{psw-2017}
\begin{align}
\vspace{-0.80cm}
T^q  & = \frac{- iC_q}{2Q} \bar{N}(p') 
   \left[  {\mathcal{H}}^{\nu} \slashed{n} 
           - \tilde {\mathcal{H}} ^{\nu} \slashed{n} \gamma^5 
   + {\mathcal E}^{\nu} \frac{i\sigma^{n\Delta}}{2m_N} 
        - \tilde {\mathcal E}^{\nu} 
        \frac{\gamma ^5 \Delta \cdot n}{2m_N}\right] N(p),
\nonumber \\[-0.10cm]
T^g & =  \frac{- i C_g}{2Q}  \bar{N}(p') \left[ {\cal H}^g \slashed{n}
      +{\cal E}^g\frac{i\sigma^{n\Delta}}{2m_N} \right] N(p) .
\label{eqn:quark-gluon-GPDs}
\end{align}
\ \vspace{-0.50cm}

\noindent
Here, charged-current reactions probe the difference between
the $u$ and $d$-quark GPDs:
$F^{\nu}(x,\xi,t) = F^d(x,\xi,t) - F^u(-x,\xi,t)$, 
where $F = H,\ \tilde H,\ E,\ \tilde E$.
The $C_q$ and $C_g$ are coupling constants with 
color and flavor factors defined as
$C_q = 2\pi C_F \alpha_s V_{dc}/3$ with 
$C_F= 4/3$, and 
$C_g = \pi T_f \alpha_s V_{du}/3$ with $T_f =1/2$.
The momentum factor $Q$ is defined as $Q^2 = -q^2$ 
by the momentum transfer $q$, 
$\slashed{n}$ is given by $\slashed{n} \equiv n_\mu \gamma^\mu = \gamma^+$
with $n^\mu = (1,\,0,\,0,\,-1)/\sqrt{2}$,
$\sigma^{n\Delta}$ is $\sigma^{n\Delta}=\sigma^{\mu\nu}n_\mu \Delta_\nu$,
and $m_N$ is the nucleon mass.
The functions ${\cal F }^{\nu}$ and ${\cal F }^{g}$ are defined
by including the pion distribution amplitude $\phi_\pi$ as
\vspace{-0.10cm}
\begin{align}
{\cal F }^{\nu} = 2 f_{\pi}\int_0^1 \frac{\phi_\pi (z)dz}{1-z}
    \int_{-1}^1 dx \frac{F^{\nu}(x,\xi,t)}{x-\xi +i\epsilon} , \ \ \ 
{\cal F }^{g} = \frac{8 f_{\pi}}{\xi} \int _0^1\frac{\phi_\pi (z)dz}{z (1- z)} 
\int _{-1}^{1}dx \frac{F^{g}(x,\xi,t)}{x-\xi +i\epsilon} ,
\label{eqn:f-quark-gluon}
\end{align} 
\ \vspace{-0.50cm}

\noindent
where $f_\pi$ is the pion decay constant.

\begin{figure}[t]
\vspace{-0.20cm}
\begin{minipage}[c]{0.48\textwidth}
\begin{center}
    \includegraphics[width=6.5cm]{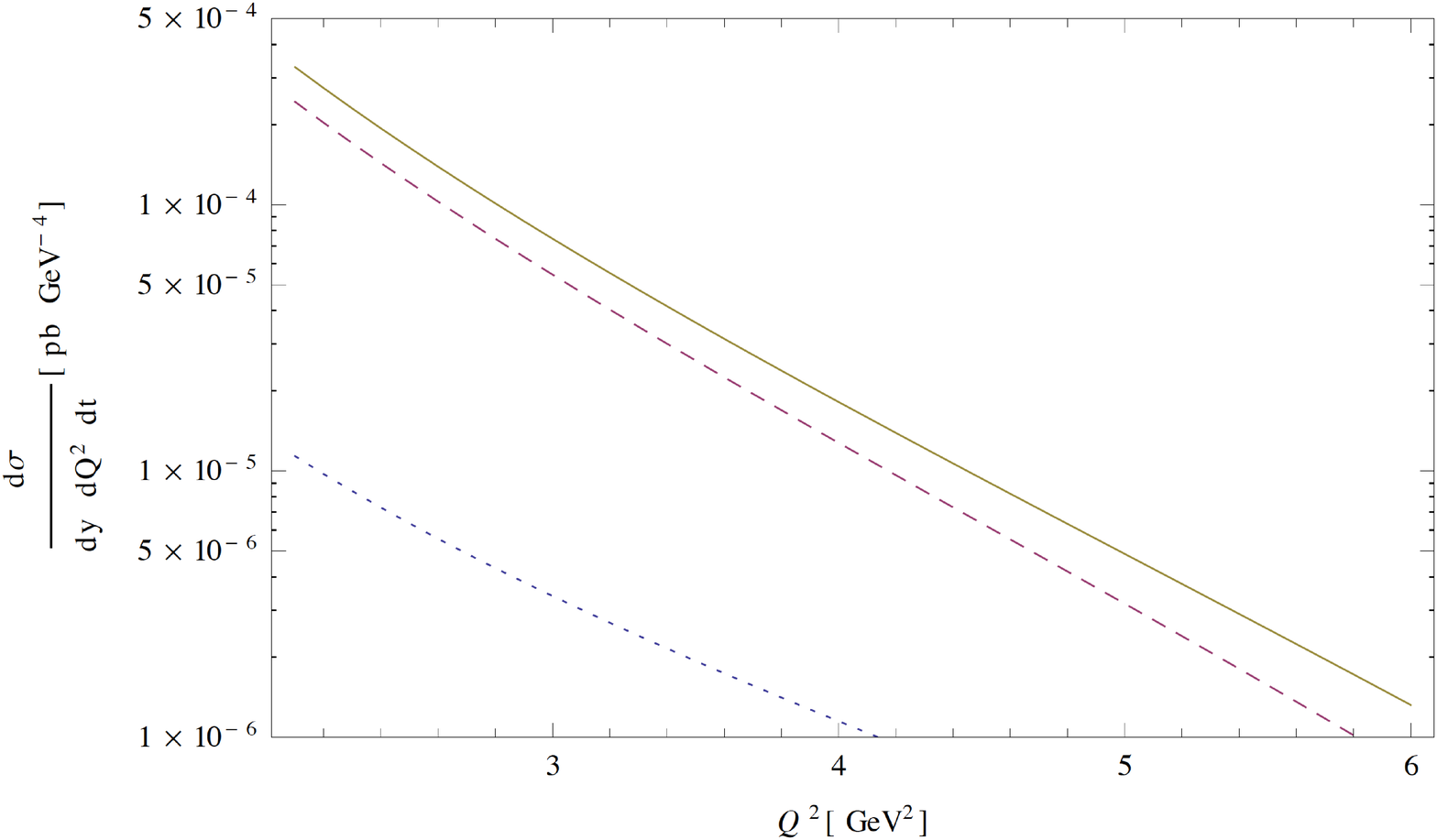}
\end{center}
\vspace{-0.60cm}
\caption{$\nu p \to \ell^- \pi^+ p$ cross section
at $y=0.7$ and $s=20$ GeV$^2$ \cite{psw-2017}. 
The dashed and dotted curves indicate gluon and quark
contributions, respectively. 
The solid curve is their summation.
}
\label{fig:pi-plus-cross}
\end{minipage}
\hspace{0.20cm}
\begin{minipage}[c]{0.48\textwidth}
\begin{center}
    \includegraphics[width=6.5cm]{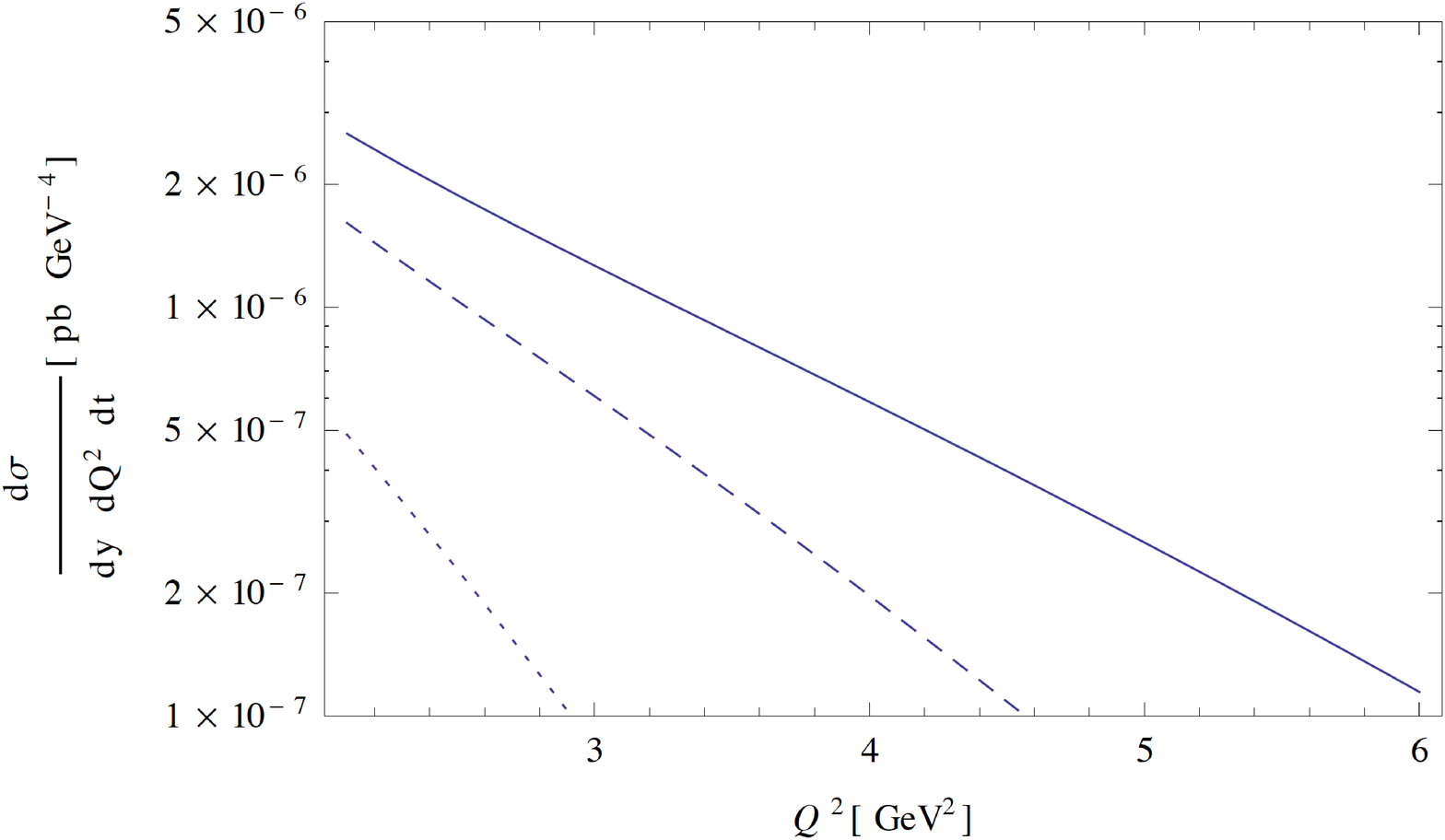}
\end{center}
\vspace{-0.60cm}
\caption{$\nu n \to \ell^- \pi^0 p$ cross section
at $s=20$ GeV$^2$ \cite{psw-2017}. 
The solid, dashed, and dotted curves indicate the corss
sections at $y=0.7$, 0.5, and 0.3, respectively. 
There is no gluon contribution for the $\pi^0$ production.
}
\label{fig:pi-0-cross}
\end{minipage}
\end{figure}

In terms of these GPDs, the cross section is written as
\vspace{-0.10cm}
\begin{align}
& \frac{d^4\sigma_{\nu N\to l^- N' \pi}}{dy\, dQ^2\, dt\,  d\varphi}
 =  \frac{G_F^2 Q^2 \, \varepsilon \sigma_{L} }
   {32 (2 \pi)^4 (1-\epsilon)(s-m_N^2)^2 y \sqrt{ 1+4x_B^2m_N^2/Q^2}} \, ,
\nonumber \\[-0.20cm]
& \ \hspace{0.4cm} 
\sigma_{L} =    \frac{1} { Q^2}\biggl\{ [\, 
      |C_q{\mathcal{H}}^{\bar q} + C_g{\mathcal{H}}^{g}|^2 
      + |C_q\tilde{\mathcal{H}}^{\bar q}|^2 ] (1-\xi^2) 
      +\frac{\xi^4}{1-\xi^2} [\,   |C_q {\mathcal{E}}^{\bar q}
      +C_g {\mathcal{E}}^{g} |^2 
      + |C_q\tilde{\mathcal{E}}^{\bar q} |^2 ]  
\nonumber  \\[-0.20cm]
& \ \hspace{2.0cm}
      -2 \xi^2 {\mathcal R}e  [C_q{\mathcal{H}}^{\bar q} 
      + C_g{\mathcal{H}}^{g}] [C_q {\mathcal{E}}^{\bar q }
      + C_g {\mathcal{E}}^{g}] ^* 
      -2 \xi^2 {\mathcal R}e 
       [C_q\tilde{\mathcal{H}}^{\bar q} ] 
       [C_q \tilde{\mathcal{E}}^{\bar q }]^* \biggr\} ,
\label{eqn:sigma-l}
\end{align}
\ \vspace{-0.70cm}

\noindent
where $y= p \cdot q / p\cdot k$, 
$Q^2 = x_B y (s-m_N^2)$, and $\varepsilon \simeq (1-y)/(1-y+y^2/2)$.
The longitudinal cross-section $\sigma_L$ is defined 
by the hadron tensor and the photon-polarization vector as
$\sigma_L=\epsilon_L^{* \mu} W_{\mu \nu} \epsilon_L^\nu$.
The obtained cross sections are shown in Figs.\,\ref{fig:pi-plus-cross}
and \ref{fig:pi-0-cross} for the $\pi^+$ and $\pi^0$ productions, 
respectively, at $s=20$ GeV$^2$.
As shown in these figures, 
both quark and gluon processes contribute to the $\pi^+$ production 
and the gluon contribution is much larger, whereas 
there is no contribution to the $\pi^0$ production
from the gluon GPD.
Therefore, the neutrino GPD measurement is valuable 
for clarifying the quark and gluon GPDs and
the flavor dependence in the quark GPDs.

\subsection{Opportunities for GPD Measurements at LBNF} 
\label{sec:neutrino-GPD} 

The future LBNF at the Fermi National Laboratory 
will deliver neutrino and antineutrino beams of unprecedented intensity 
with broad energy spectra.  In addition to the default beam optimized 
for long-baseline oscillation measurements in the energy range 0.5-5 GeV, 
a higher energy option (mostly in the 2-15GeV range) is possible 
for precision measurements and searches for new physics 
beyond the Standard Model.
The near detector complex of the Deep Underground Neutrino Experiment (DUNE) 
will include a high resolution on-axis detector which can address some 
of main limitations of (anti)neutrino experiments providing 
an accurate control of the targets and fluxes~\cite{Petti:2019asx}. 
In particular, it will allow precision measurements of $\nu$ and $\bar \nu$ 
interactions on both hydrogen (H) and various nuclear targets (A) 
in combination with the high intensity and the energy spectra 
of the LBNF beams. 
The kinematic coverage is dominated by inelastic interactions 
-- more than 54\% of the events with the default low energy beam 
and most of the events with the high energy option have $W>1.4$ GeV 
-- offering a good sensitivity to the GPD measurements 
via the pion production processes $\nu(\bar \nu) N \to l N^\prime \pi$. 
The availability of a free proton target H will give access 
to high statistics measurements of the following channels: 
(a) $\nu_\mu p \to \mu^- \pi^+ p$; (b) $\bar \nu_\mu p \to \mu^+ \pi^- p$; 
(c) $\bar \nu_\mu p \to \mu^+ \pi^0 n$. 
While both quark and gluon GPDs contribute to the $\pi^\pm$ production, 
no gluon contribution is present for the $\pi^0$ production. 
The last two measurements with also provide information about 
the GPDs in a free neutron target since it is expected 
$\sigma (\bar \nu_\mu p \to \mu^+ \pi^- p) 
 = \sigma(\nu_\mu n \to \mu^- \pi^+ n)$ and 
$\sigma (\bar \nu_\mu p \to \mu^+ \pi^0 n) 
= \sigma (\nu_\mu n \to \mu^- \pi^0 p)$~\cite{psw-2017}. 
The study of the flavor dependence of the GPDs in free nucleons 
will be complemented by similar measurements performed 
simultaneously on a variety of nuclear targets (C, Ar, etc.) 
within the same detector. A comparison between measurements 
on H and on the nuclear targets can provide valuable information 
about the nuclear modifications of the GPDs. 
The nuclear targets will also extend the study of the flavor dependence 
of the GPDs by giving access to additional channels: 
(a) $\nu_\mu n \to \mu^- \pi^+ n$; (b) $\nu_\mu n \to \mu^- \pi^0 p$; 
(c) $\bar \nu_\mu n \to \mu^+ \pi^- n$. 
A sizable statistics for the various single pion production channels 
is expected to be collected from both H and nuclear targets with 
the default low energy LBNF beam~\cite{Petti:2019asx,Duyang:2018lpe}.
The high energy beam options will significantly enhance the sensitivity 
to the GPD measurements increasing the kinematic 
overlap with complementary EIC measurements.

\section{Summary}
\label{summary}

The GPD studies will be crucial in understanding the origins of hadron
spins and masses in terms of quarks and gluons. 
The spacelike GPDs are measured in charged-lepton scattering processes,
deeply virtual Compton scattering and meson productions,
and timelike GPDs are investigated by two-photon processes.
Using the LBNF neutrino beam, we can access the spacelike GPDs 
in neutrino reactions.
As the neutrino DIS measurements played an important role in 
establishing the flavor-dependence of the PDFs and the valence-quark 
distribution functions, the neutrino GPD measurements should be 
complementary to the charged-lepton ones.
The high resolution on-axis detector in the DUNE near detector complex 
will be capable of detailed GPD studies on both free protons 
and nuclei in future, providing insights on the hadron spins and masses.

\section*{Acknowledgments}

Figures 5 and 6 are used with the copyright permission of American
Physical Society and authors.
S.K. was partially supported by 
Japan Society for the Promotion of Science (JSPS) Grants-in-Aid 
for Scientific Research (KAKENHI) Grant Number 19K03830.
R.P. was supported by grant DE-SC0010073 from the Department of Energy, USA.



\end{document}